\documentclass[runningheads]{llncs}
\usepackage{graphicx}
\usepackage{amsmath}
\usepackage{amssymb}
\usepackage{caption}
\usepackage{wrapfig}
\usepackage{hyperref}
\usepackage{pifont}
\newcommand{\cmark}{\ding{51}}%
\newcommand{\xmark}{\ding{55}}%
\newcommand{\printfnsymbol}[1]{%
  \textsuperscript{\@*}%
  }

\usepackage{booktabs}
\begin{document}
%
\title{`Project \& Excite' Modules for Segmentation of Volumetric Medical Scans }
\titlerunning{`Project \& Excite'}
\author{Anne-Marie Rickmann \inst{1,2}\thanks{The authors contributed equally.} \and Abhijit Guha Roy\inst{1,2}\printfnsymbol{1} \and
Ignacio Sarasua\inst{1}\printfnsymbol{1} \and
Nassir Navab \inst{2,3} \and
Christian Wachinger \inst{1} }
%
%

\authorrunning{Rickmann et al.}
%
\institute{
Artificial Intelligence in Medical Imaging (AI-Med), KJP, LMU M\"unchen, Germany \and
Computer Aided Medical Procedures, Technische Universit\"at M\"unchen, Germany
\and
Computer Aided Medical Procedures, Johns Hopkins University, USA}
\maketitle              
\begin{abstract}
Fully Convolutional Neural Networks (F-CNNs) achieve state-of-the-art  performance for image segmentation in medical imaging.  
Recently, squeeze and excitation (SE) modules and variations thereof have been introduced to recalibrate feature maps channel- and spatial-wise, which can boost performance while only minimally increasing model complexity. 
So far, the development of SE has focused on 2D images. 
In this paper, we propose `Project \& Excite' (PE) modules that base upon the ideas of SE and extend them to operating on 3D volumetric images. 
`Project \& Excite' does not perform global average pooling, but squeezes feature maps along different slices of a tensor separately to retain more spatial information that is subsequently used in the excitation step.
We demonstrate that PE modules can be easily integrated in 3D U-Net, boosting performance by $5\%$ Dice points, while only increasing the model complexity by $2\%$.
We evaluate the PE module on two challenging tasks, whole-brain segmentation of MRI scans and whole-body segmentation of CT scans. Code: \href{https://github.com/ai-med/squeeze_and_excitation}{https://github.com/ai-med/squeeze\_and\_excitation} 

\end{abstract}
\section{Introduction}
Fully convolutional neural networks (F-CNNs) have been widely adopted for semantic image segmentation in computer vision~\cite{long2015fully} and medical imaging~\cite{ronneberger2015u}.
As computer vision tasks mainly deal with 2D natural images, most of the architectural innovations have focused towards 2D CNNs.
These innovations are often not applicable for processing volumetric medical scans like CT, MRI and PET.
For segmentation, 2D F-CNNs were used to segment 3D medical scans slice-wise. In such an approach the contextual information from adjacent slices remains unexplored, which might lead to imperfect segmentations, especially if the target class is small. 
Hence, the natural choice of segmenting 3D scans would be to use 3D F-CNN architectures.
However, there exist some practical challenges in using 3D F-CNNs: 
(i) 3D F-CNNs require large amount of GPU RAM space for training, and 
(ii) the number of weight parameters are much higher than for its 2D counter-part, which can make the models prone to over-fitting with limited training data.
Although the first issue can be effectively addressed using recent GPU clusters, the second issue still remains.
This problem is prominent in medical applications, where training data is commonly very limited. Most datasets often contain only about 15-20 annotated training scans. 
To overcome the problem of over-fitting, 3D F-CNNs are carefully engineered for a task to minimize the model complexity by reducing the number of convolutional layers or by decreasing the number of channels per convolutional layer. 
Although this might aid training models with limited data, the exploratory capacity of the 3D F-CNN gets limited.
In such a scenario, it is necessary to ensure that the learnable parameters within the F-CNN are maximally utilized to solve the task at hand.
Recently, a computational module termed `Squeeze and Excite' (SE) block~\cite{Hu_2018_CVPR} has been introduced to recalibrate CNN feature maps, which boosts the performance while increasing model complexity marginally. 
This is performed by modeling the interdependencies between the channels of feature maps, and learning to provide attention on specific channels depending on the task. 
This idea was also extended to medical image segmentation~\cite{roy2019recalibrating}, where it was demonstrated that such light-weight blocks can be a better architectural choice than extra convolutional layers.
Although, SE blocks were customarily designed for 2D architectures, they have recently been extended to 3D F-CNNS to aid volumetric segmentation~\cite{zhu2019anatomynet}.

\begin{wrapfigure}{r}{0.55\textwidth}
    \centering
    \vspace{-.7cm}
     \includegraphics[width= 0.48 \textwidth]{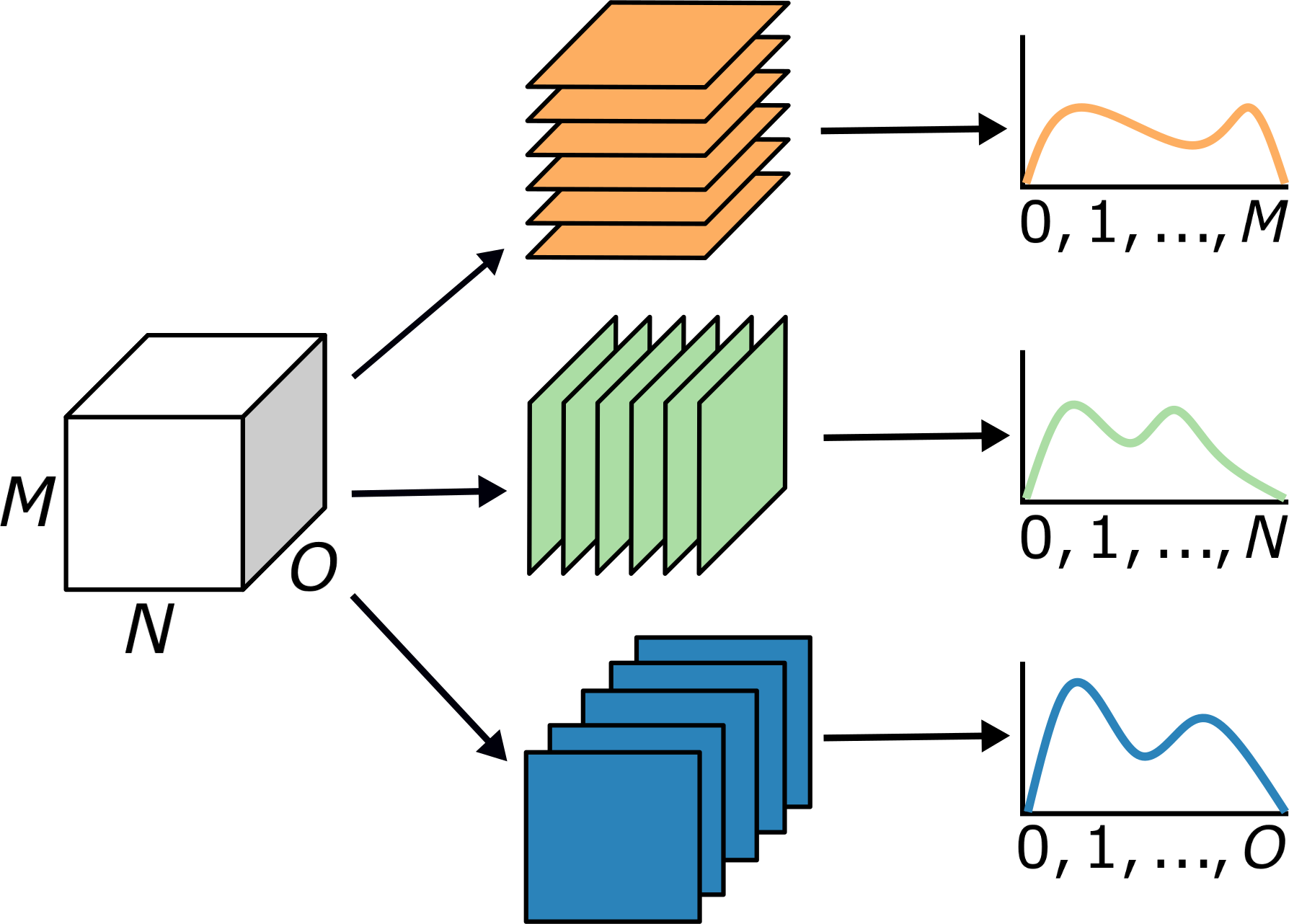}
    \caption{Projections in `PE' block}
    \vspace{-.7cm}
    \label{fig:motivation}
\end{wrapfigure}

In this paper, we propose the `Project \& Excite' (PE) module, a new computational block custom-made to recalibrate 3D F-CNNs.
Zhu et al.~\cite{zhu2019anatomynet} directly extended the concept of SE to 3D by averaging the 4D tensor over all spatial dimensions to generate a channel descriptor for recalibration.
We hypothesize that removing all spatial information leads to a loss of relevant information, particularly for segmentation, where we need to exactly localize anatomical structures. 
In contrast, we aim at preserving the spatial information without any excess model complexity or FLOP operations, which is relevant for fine-grained volumetric segmentation.
We draw our inspiration from traditional tensor slicing techniques, by averaging along the three principle axes of the tensor as indicated in Fig.~\ref{fig:motivation}. We term this operation the `Projection' operation.
By this, we get three projection-vectors indicating the relevance of the slices along the three axes. A spatial location is important if all the corresponding slices associated with it provide higher estimates.
So, instead of learning the dependencies of the scalar values across the channels as in~\cite{zhu2019anatomynet}, we learn the dependencies of these projection-vectors across the channels for excitation.
Also, PE blocks provide a global receptive field to the network at every stage.

Our contributions are: (i) we propose a new computational block termed `Project \& Excite' for recalibration of 3D F-CNNs,
(ii) we demonstrate that our proposed PE blocks can easily be integrated into any F-CNNs boosting the segmentation performance, especially for small target classes,
(iii) we demonstrate that PE blocks minimally increase the model complexity in contrast to using more convolutional layers, while providing much higher segmentation accuracy, substantiating its effectiveness in recalibration.

\section{Methods}
`Squeeze \& excite' (SE) blocks $\mathbf{F}_{se}(\cdot)$ take a feature map $\mathbf{U}$ as input and recalibrate it to $\mathbf{\hat{U}} = \mathbf{F}_{se}(\mathbf{U})$. 
Let $\mathbf{\hat{U}} \in \mathbb{R}^{H\times W \times D \times C}$, with height $H$, width $W$, depth $D$, and number of channels $C$.
Commonly, SE blocks are placed after every encoder and decoder blocks of an F-CNN. 
In this section, we detail the extension of SE to 3D F-CNNs and our proposed `Project \& Excite' blocks. 

\vspace{2mm}
\noindent
\textbf{3D `Squeeze \& Excite' Module: }
This 3D SE block~\cite{zhu2019anatomynet}, that can be termed channel SE (cSE) module, is a direct extension of the 2D SE blocks proposed in~\cite{Hu_2018_CVPR} to a 3D version.  
The transformation $\mathbf{F}_{se}(\cdot)$ is divided into the squeeze operation $\mathbf{F}_{sq}(\cdot)$ and excite operation $\mathbf{F}_{ex}(\cdot)$.
The squeeze operation $\mathbf{F}_{sq}(\cdot)$ performs a global average pooling operation that squeezes the spatial content of the input $\mathbf{U}$  into a scalar value per channel $\mathbf{z} \in \mathbb{R}^{C}$.
The excitation operation $\mathbf{F}_{ex}(\cdot)$ takes in $\mathbf{z}$ and adaptively learns the inter-channel dependencies by using two fully-connected layers. The operations are defined as:
\begin{align}
    z_c &= \mathbf{F}_{sq}(\mathbf{u}_c) = \frac{1}{H}\frac{1}{W}\frac{1}{D} \sum_{i=1}^{H}\sum_{j=1}^W\sum_{k=1}^D \mathbf{u}_c(i,j,k), \\
    \mathbf{\hat{z}} &= \mathbf{F}_{ex}(\mathbf{z},\mathbf{W}) =  \sigma(\mathbf{W}_2\delta(\mathbf{W}_1 \mathbf{z})),
\end{align}

\noindent
with $\delta$ denoting the ReLU nonlinearity, $\sigma$ the sigmoid layer, $\mathbf{W}_1 \in \mathbb{R}^{\frac{C}{r}\times C}$ and $\mathbf{W}_2 \in \mathbb{R}^{C \times \frac{C}{r}}$ the weights of the fully-connected layers and $r$ is the channel reduction factor similar to~\cite{Hu_2018_CVPR}.
The output of the 3D cSE module is defined by a channel-wise multiplication of $\mathbf{U}$ with $\mathbf{\hat{z}}$.
The $c^{\mathrm{th}}$ channel of $\mathbf{\hat{U}}$ is defined as: $ \mathbf{\hat{u}}_c = \mathbf{F}_{ex}(\mathbf{F}_{sq}(\mathbf{u}_c)) \mathbf{u}_c = \hat{z}_c \mathbf{u}_c$.

\begin{figure}[h]
  \centering
  \includegraphics[width=0.8\linewidth]{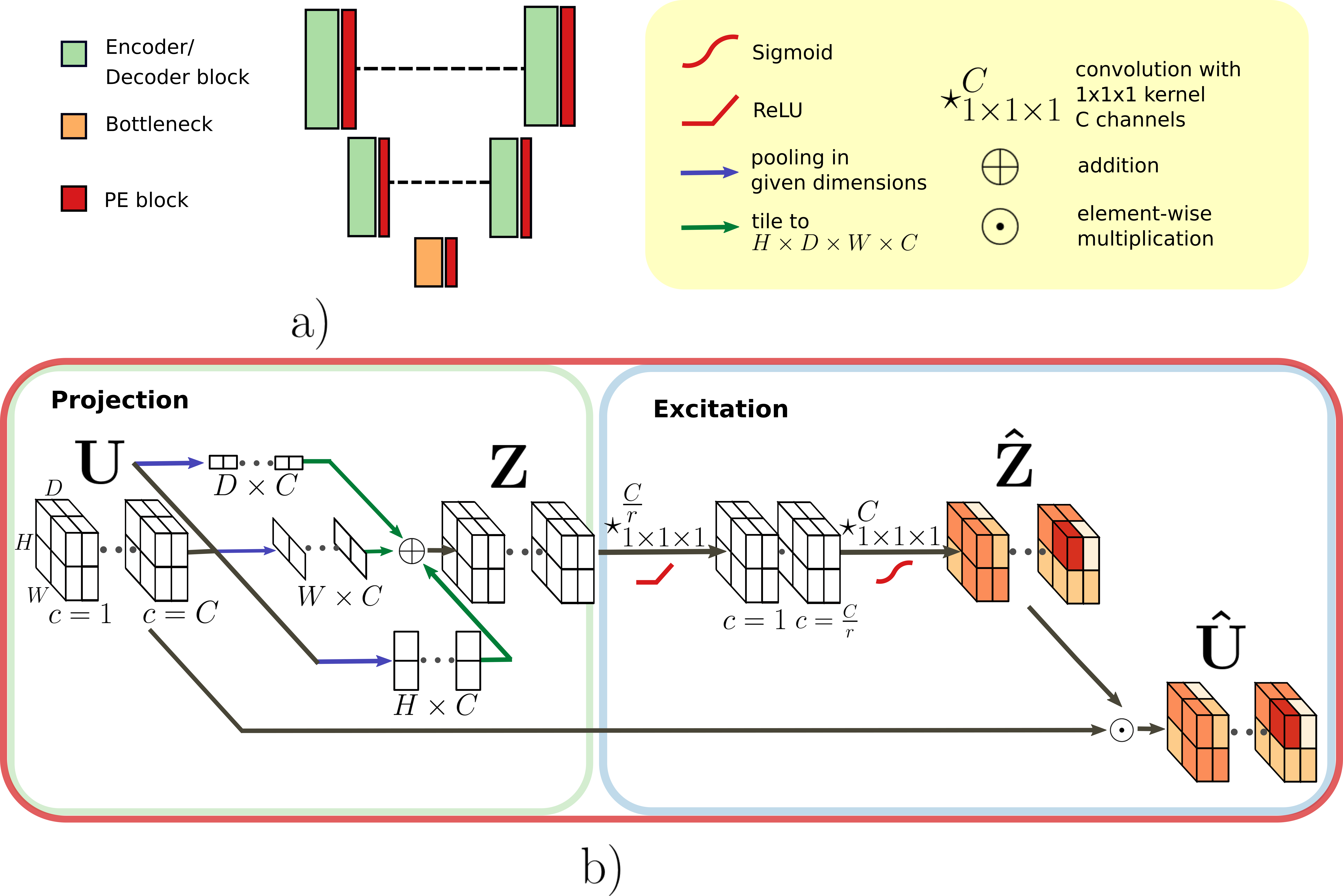}
  \caption{a): typical encoder/decoder based F-CNN architecture with PE blocks placed after each block.
  b): Illustration of the proposed 'Project\& Excite' block. Projection operation, with the 3 different pooling operations and Excitation operation with 2 convolutional layers and recalibration of the feature map.}
\label{fig:pe_block}
\end{figure}

\vspace{2mm}
\noindent
\textbf{3D `Project \& Excite' Module: }
The 3D cSE module squeezes spatial information of a volumetric feature map into one scalar value per channel. Especially in the first/last layers of a typical architecture, these feature maps have a high spatial extent. Our hypothesis is that a volumetric input of large size holds relevant spatial information which might not be properly captured by a global pooling operation. Hence, we introduce the `Project \& Excite' module that retains more of the valuable spatial information within our proposed \emph{projection operation instead of spatial squeeze operation}. This follows the excite operation, which learns inter-dependencies between the projections across the different channels. Thus, it combines spatial and channel context for recalibration. 
The architectural details of the `PE' block is illustrated in Fig.~\ref{fig:pe_block}.

The projection operation $\mathbf{F}_{pr}(\cdot)$ is separated into three projection operations ($\mathbf{F}_{{pr}_H}(\cdot)$, $\mathbf{F}_{{pr}_W}(\cdot)$, $\mathbf{F}_{{pr}_D}(\cdot)$) along the spatial dimensions with outputs $\mathbf{z}_{h_c} \in \mathbb{R}^{C \times H}$, $\mathbf{z}_{w_c} \in \mathbb{R}^{C \times W}$ and $\mathbf{z}_{d_c} \in \mathbb{R}^{C \times D}$. The projection operations are done by average pooling defined as:
\begin{align}
    \mathbf{z}_{h_c}(i) &= \mathbf{F}_{pr_H}(\mathbf{u}_c) = \frac{1}{W}\frac{1}{D}\sum_{j=1}^W\sum_{k=1}^D \mathbf{u}_c(i,j,k), \quad i \in \{1, \ldots, H\}\\
    \mathbf{z}_{w_c}(j) &= \mathbf{F}_{pr_W}(\mathbf{u}_c) = \frac{1}{H}\frac{1}{D} \sum_{i=1}^{H}\sum_{k=1}^D \mathbf{u}_c(i,j,k), \quad j \in \{1, \ldots, W\}\\
    \mathbf{z}_{d_c}(k) &= \mathbf{F}_{pr_D}(\mathbf{u}_c) = \frac{1}{H}\frac{1}{W} \sum_{i=1}^{H}\sum_{j=1}^W \mathbf{u}_c(i,j,k), \quad k \in \{1, \ldots, D\}.
\end{align}
The outputs $\mathbf{z}_c$ are tiled to the shape $H \times W \times D \times C$ and added to obtain $\mathbf{Z}$, which is then fed to the excitation operation $\mathbf{F}_{ex}(\cdot)$, which is defined by two convolutional layers followed by a ReLU and sigmoid activation respectively. The convolutional layers have kernel size $1 \times 1 \times 1$, to aid modelling of channel dependencies. The first layer reduces the number of channels by $r$, and the second layer brings the channel dimension back to the original size. The excite operation is defined as:
\begin{equation}
    \hat{\mathbf{U}} = \hat{\mathbf{Z}} \odot \mathbf{U} = \mathbf{F}_{ex}(\mathbf{Z}) \odot \mathbf{U} = \sigma (\mathbf{V}_2 \star \delta (\mathbf{V}_1 \star \mathbf{Z})) \odot \mathbf{U},
\end{equation}
where $\star$ describes the convolution operation, $\odot$ indicates point-wise multiplication, $\mathbf{V}_1 \in \mathbb{R}^{1 \times 1 \times 1 \times \frac{C}{r}}$ and $\mathbf{V}_2 \in \mathbb{R}^{1 \times 1 \times 1 \times C}$ the convolution weights, $\sigma$ the sigmoid and $\delta$ the ReLU activation function.
The final output of the PE block $\hat{\mathbf{U}}$ is obtained by an element-wise multiplication of the feature map $\mathbf{U}$ and $\hat{\mathbf{Z}}$.

\vspace{-2mm}
\section{Experimental Setup}
\noindent
\label{sec:exp}
\textbf{Datasets: } For evaluation, we choose two challenging 3D segmentation tasks.
(i) \emph{Whole-brain segmentation of MRI T1 scans:} For this task, we use the Multi-Atlas Labelling Challenge (MALC) dataset~\cite{landman2012miccai}. 
It consists of $30$ T1 MRI volumes of the brain. We segment the brain volumes into 32 cortical and subcortical structures. $15$ scans were used for training, $3$ scans for validation and the remaining $12$ scans for testing. Manual segmentations for MALC were provided by Neuromorphometrics, Inc.
(ii) \emph{Whole-body segmentation of contrast enhanced CT scans:} For this task, we use the Visceral dataset~\cite{visceral}. The gold corpus of the dataset has $20$ annotated scans. We perform 5-fold cross-validation. One scan from the test fold was kept as validation set. We segment $14$ organs from thorax and abdomen. 
Both datasets have common challenges w.r.t the limited amount of training scans and severe class-imbalance across the target classes.

\noindent
\textbf{Training Setup: } We choose 3D U-Net~\cite{cciccek20163d} architecture for our experimental purposes. 
Instead of using 3D sub-volumes, we train with whole 3D scans, for which we slightly modified the 3D U-Net architecture to ensure proper trainability. 
Our design consists of 3 encoder and 3 decoder blocks, with only the first two encoders performing downsampling, and the last two decoders performing upsampling. 
Each encoder/decoder consists of 2 convolutional layers with kernel size of $3\times3\times3$. Further, the number of output channels at every encoder/decoder block was reduced to half of original size used in 3D U-Net to keep the model complexity low. For example, the two convolutions in encoder 1 have number of channels $\{ 16, 32 \}$ instead of $\{ 32, 64 \}$ and so on.
We performed preliminary experiments to conclude that this architecture was the best for our application.

\noindent
\textbf{Training Parameters: } Due to the large and variable dimensions of the input volumes we chose a batch size of 1 for training purpose. Also, this configuration totally occupied the $2\times12$ GB RAM of the TITAN Xp GPU. 
As low batch sizes make training unstable with Batch normalization layers, we use Instance normalization~\cite{instancenorm} instead which is agnostic to batch size.
Optimization was done using SGD with momentum of 0.9. The learning rate was initially set to 0.1 and was reduced by a factor of 10 when validation loss plateaued.
On-the-fly data augmentation using elastic deformations and random rotations was performed on the training set.
We used a combined Cross Entropy and Dice loss with the Cross Entropy loss being weighted using median frequency balancing to tackle the high class imbalance, similar to~\cite{roy2017error}.

\section{Experimental Results and Discussions}
\label{sec:res}

\noindent
\textbf{Position of `PE' blocks: } In this section, we investigate the positions at which our proposed `Project \& Excite' (PE) blocks need to be placed within the 3D U-Net architecture. We explored $6$ possibilities by placing them after every encoder block (P1), after every decoder block (P2), after the bottleneck block (P3), after both encoder and decoder blocks (P4), after each encoder block and bottleneck (P5), and finally after all the blocks (P6). We present the results of all these
\begin{wraptable}{r}{0.6\textwidth}
    \centering
    \scriptsize
    \caption{Mean Dice score on MALC dataset due to placement of `PE' blocks within 3D U-Net architecture.}
    \begin{tabular}{l c c c c}
        \toprule
        & \multicolumn{3}{c}{Position of `PE' block}&\\
         & Encoders & Bottleneck & Decoders & Mean Dice $\pm$ std \\
        \midrule
         3D U-Net & \xmark  & \xmark & \xmark & $0.802 \pm 0.171$\\ \hline
         P1 & \cmark & \xmark & \xmark & $0.828 \pm 0.111$\\
         P2 & \xmark & \xmark & \cmark & $0.796 \pm 0.215$\\
         P3 & \xmark & \cmark & \xmark & $0.822 \pm 0.144$\\
         P4 & \cmark & \xmark & \cmark & $0.819 \pm 0.159$\\
         P5 & \cmark & \cmark & \xmark & $0.818 \pm 0.156$\\
         P6 & \cmark & \cmark & \cmark & $ \mathbf{0.843} \pm 0.079$\\
        \bottomrule
    \end{tabular}
        \vspace{-.5cm}
    \label{tab:placement_of_se}
\end{wraptable}
configurations for MALC dataset in Tab.~\ref{tab:placement_of_se} and compared against having no `PE' blocks. We observed that placing the blocks after every encoder, decoder and bottleneck provided the best accuracy, boosting by $4\%$ Dice points. Also, we observed that placing it after encoder and bottleneck blocks improves the Dice score by $2\%$ points, whereas placing it after decoder blocks does not effect the performance. We conclude that `PE' blocks are most effective in encoder and bottleneck positions of F-CNN. In the following experiments, we place the `PE' blocks after every encoder, decoder and bottleneck blocks.

\noindent
\textbf{Model Complexity: }
Here we investigate the increase of model complexity due to addition of `PE' blocks within 3D U-Net architecture. We compare the PE blocks with 3D cSE blocks complexity-wise and report them in Tab.~\ref{tab:complexity}. We present results on MALC dataset. We observe that both PE blocks and cSE blocks cause the same fraction of $1.97\%$ increase in model complexity, whereas PE blocks provide a $2\%$ higher boost in performance at the same expense. One might think that this boost in performance is due to the added complexity, which might also be gained by adding more convolutional layers. 
\begin{wraptable}{r}{0.4\textwidth}
    \centering
    \scriptsize
    \caption{Mean Dice vs model complexity measured in number of trainable parameters}
    \begin{tabular}{l c c }
        \toprule
        & Dice & Complexity\\
        \midrule
        3D-Unet \cite{cciccek20163d} & $0.802$ & $5.57 \cdot 10^6$\\
        \midrule
        + 3D cSE \cite{zhu2019anatomynet} & $0.825$ &  $ + 1.97 \%$\\
        + PE & $\mathbf{0.843}$ & $ + 1.97 \%$\\
        + Encoder/Decoder & $0.779$ & $+ 39.7 \%$\\
        + 2 conv layers & $0.826$ & $+ 3.97 \%$\\
        \bottomrule
    \end{tabular}
    \vspace{-.5cm}
    \label{tab:complexity}
\end{wraptable}
We investigated this matter by conducting two more experiments. First, we added an extra encoder and decoder block within the architecture. This immensely increased the model complexity by almost $40\%$ and we observed a drop in Dice performance. One possible reason might be due to over-fitting given the limited data samples and sudden increase in model complexity. So, next we only added two additional convolutional layers at the second encoder and second decoder to make sure that the increase in model complexity is only marginal ($\sim4\%$), not risking over-fitting. Here, we did observe a boost in performance similar to cSE with double the increase in parameters, but still failed to match the performance of our PE blocks. Thus, we can conclude that PE blocks are in fact more effective than simply adding convolutional layers.

\begin{table}[h]
    \centering
    \scriptsize
    \caption{Comparison of 3D U-Net with 3D cSE and our proposed PE block. Mean Dice scores for selected classes of MALC and Visceral datasets. In the top table WM stands for white matter and GM for grey matter. In the bottom table L. stands for left and R. for right.}
    \label{tab3}
    \begin{tabular}{l c c c c c c}
        \toprule
         & \multicolumn{6}{c}{\bfseries MALC dataset} \\
          & Mean Dice $\pm$ std & WM & GM & Inf. Lat. Vent. & Amygdala & Accumbens\\ 
          \midrule
         3D U-Net \cite{cciccek20163d} & $ 0.802 \pm 0.171$ &$0.906$ & $0.887$&  $0.242$ & $0.761$ & $0.483$ \\
         3D cSE \cite{Hu_2018_CVPR,zhu2019anatomynet} & $0.825 \pm 0.119$ & $0.907$ & $0.888$ & $0.403$ &  $0.761$ & $0.704$\\
         Project \& Excite & $ \mathbf{0.843} \pm 0.079$ & $\mathbf{0.916}$ & $ \mathbf{0.899}$ & $ \mathbf{0.604}$ & $\mathbf{0.789}$ & $\mathbf{0.735}$\\
         \midrule
        & \multicolumn{6}{c}{\bfseries Visceral dataset} \\
         &Mean Dice $\pm$ std & Liver & R. Lung & R. Kidney & Trachea & Sternum  \\
         \midrule
        3D U-Net \cite{cciccek20163d} & $0.810 \pm 0.137$ & $0.922$ & $0.965$ & $0.907$ & $0.815$ & $0.438$\\
        3D cSE  \cite{Hu_2018_CVPR,zhu2019anatomynet} & $0.797 \pm 0.168$ & $0.930$ & $0.966$ & $0.919$ & $0.491$ & $0.427$\\
        Project \& Excite &  $\mathbf{0.846} \pm 0.095$ & $0.931$ & $0.966$ & $ \mathbf{0.929}$ & $\mathbf{0.845}$ & $\mathbf{0.699}$\\
        \bottomrule
    \end{tabular}
    \vspace{-.5cm}
    \label{tab:main_results}
\end{table}

\begin{figure}[h]
    \centering
     \includegraphics[width=0.85\linewidth]{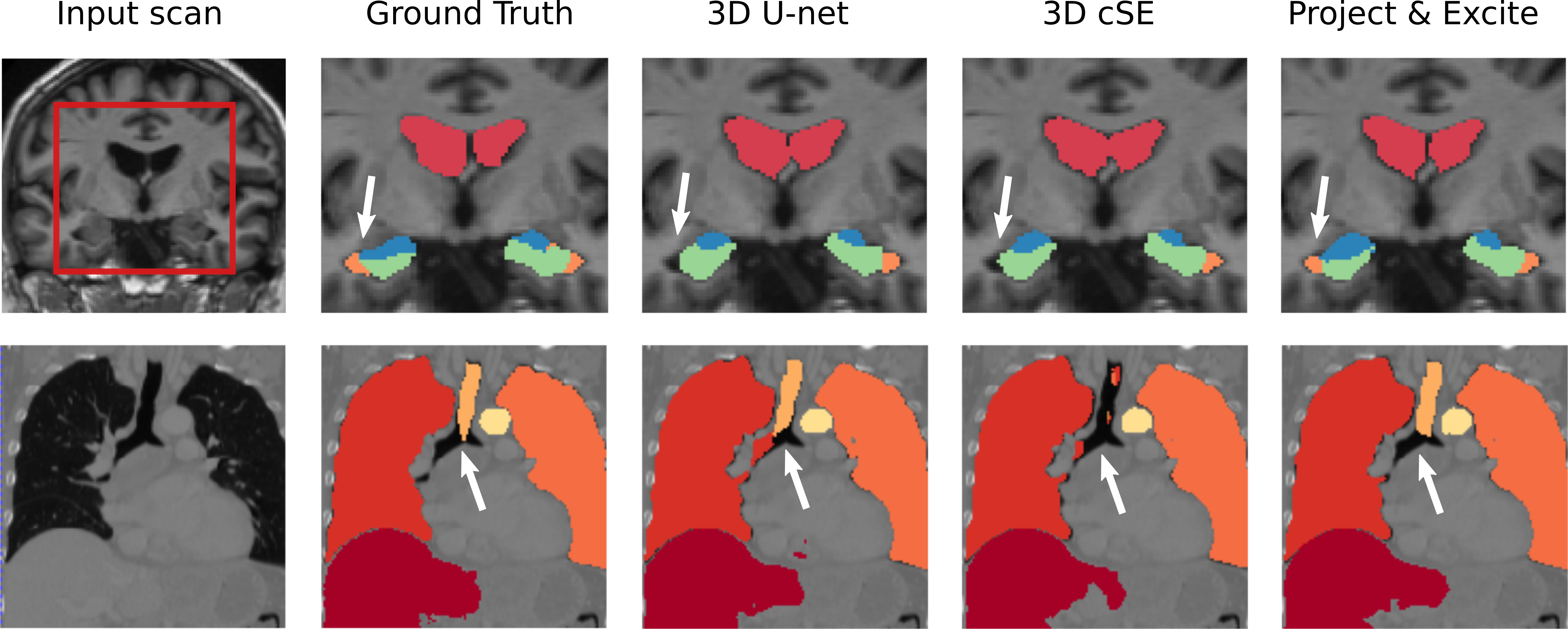}
    \caption{Input scans, manual segmentation and results for 3D U-Net, 3D cSE and our PE model, for MALC (top row) and Visceral (bottom row) datasets. Red box and white arrows point to the structures where our PE model improved the performance.}
    \label{fig:segmentations}
\end{figure}

\noindent
\textbf{Segmentation Results: }
We present the results of whole-brain segmentation and whole-body segmentation in Tab.~\ref{tab:main_results}.
We compared `PE' blocks to the 3D channel SE (cSE) blocks~\cite{zhu2019anatomynet} and the baseline 3D U-Net. The placement of the cSE blocks in the architecture was kept identical to ours.
For brain segmentation, we observe the overall mean Dice score by using 3D cSE increases by $2\%$ Dice points, whereas our proposed `PE' blocks lead to an increase of $4\%$ Dice points, substantiating its efficacy.
For whole body segmentation, the mean Dice score by using 3D cSE even decreases by $1\%$, while, when using PE blocks, it increases by $3.5\%$.
Further, we explored the impact of PE blocks on some selected structures. 
Firstly, we selected bigger structures, white and grey matter for brain segmentation, and liver and right lung for whole-body segmentation.
The boost in Dice score for white and grey matter was very marginal by using either cSE or PE blocks ranging within $1\%$ Dice points. For liver and right lung the performance using cSE or PE blocks is comparable to the baseline 3D U-Net.
Next, we analyze some smaller structures, namely inferior lateral ventricles, amygdala and accumbens for brain segmentation, and right kidney, trachea and sternum for whole body segmentation, which are difficult to segment.
We observe an immense boost in performance using PE blocks in these structures ranging between $3-36\%$ Dice points, while using cSE blocks even leads to decreasing performance for trachea and sternum.
In Fig.~\ref{fig:segmentations}, we present visualizations of the segmentation performance of PE models in comparison to baseline 3D U-Net and 3D cSE models. 
In the top row, white arrows indicate the region of left inferior lateral ventricle, which was missed by both 3D U-Net and 3D csE models. 
Our proposed PE model, however, was able to segment this very small structure.
In the bottom row, white arrows point to the bifurcation of the trachea, where the 3D U-Net is oversegmenting the right lung and 3D cSE model is missing the trachea completely.
In conclusion, we observed similar trends in both, whole-brain and whole-body segmentation, demonstrating the efficacy of PE blocks for segmentation of small structures in 3D scans.

\section{Conclusion}
We propose `Project \& Excite', a light-weight recalibration module that can be easily integrated within any 3D F-CNN architectures and boosts segmentation performance while increasing model complexity by a small fraction.  
We demonstrated that PE blocks can be an attractive alternative to adding more convolutional layers in 3D F-CNNs, especially in situations where training data and GPU resource is limited. 
We exhibited the effectiveness of `PE' blocks by conducting experiments on two challenging tasks of whole-brain and whole-body segmentation. 

\section*{Acknowledgements}
This research was partially supported by the Bavarian State Ministry of Education,
Science and the Arts in the framework of the Centre Digitisation.Bavaria (ZD.B). We thank NVIDIA corporation for GPU donation.

\vspace{-0.3cm}
\bibliographystyle{splncs04}
\bibliography{references}

\end{document}